\documentclass[trackchanges,twocolumn]{aastex701}
\usepackage{float}
\usepackage{amsmath}
\usepackage{bm}
\usepackage{rotating}
\usepackage{multirow}
\usepackage{makecell}
\usepackage{subfigure}
\usepackage{enumitem}

\begin{document}

\title{An Opacity-Free Test of the Cosmic Distance Duality Relation Using Strongly Lensed Gravitational Wave Signals with Space-Based Detector Networks}

\author[orcid=0009-0002-9540-9230]{Yong Yuan}
\affiliation{Center for Gravitational Wave Experiment, National Microgravity Laboratory, Institute of Mechanics, Chinese Academy of Sciences, Beijing, China}
\email{yuanyong@imech.ac.cn}  

\author[orcid=0000-0003-2155-3280]{Minghui Du} 
\affiliation{Center for Gravitational Wave Experiment, National Microgravity Laboratory, Institute of Mechanics, Chinese Academy of Sciences, Beijing, China}
\email{duminghui@imech.ac.cn}

\author[orcid=0000-0003-1511-5567]{Benyang Zhu}
\affiliation{Key Laboratory of Dark Matter and Space Astronomy, Purple Mountain Observatory, Chinese Academy of Sciences, Nanjing 210023, China}
\affiliation{School of Astronomy and Space Science, University of Science and Technology of China, Hefei 230026, China}
\email{byzhu@pmo.ac.cn}

\author[orcid=0009-0001-8476-7115]{Xin-yi Lin}
\affiliation{Department of Astronomy, Beijing Normal University, Beijing 100875, China}
\email{xinyilin@bnu.edu.cn}

% \author[orcid=0000-0002-7304-2426]{Huan Zhou}
% \affiliation{School of Physics and Optoelectronic Engineering, Yangtze University, Jingzhou, 434023, China}
% \email{525040@yangtzeu.edu.cn}

\author[orcid=0000-0001-5033-6168, gname=Wen-Fan, sname=Feng]{Wen-Fan Feng}
\affiliation{Kavli Institute for Astronomy and Astrophysics, Peking University,
Beijing 100871, China}
\email{fengwf@pku.edu.cn}  

\author[orcid=0000-0002-3543-7777]{Peng Xu}
\affiliation{Center for Gravitational Wave Experiment, National Microgravity Laboratory, Institute of Mechanics, Chinese Academy of Sciences, Beijing, China}
\affiliation{Taiji Laboratory for Gravitational Wave Universe (Beijing/Hangzhou), University of Chinese Academy of Sciences, Beijing, 100049,
China}
\affiliation{Hangzhou Institute for Advanced Study, University of Chinese Academy of Sciences, Hangzhou, 310124, China}
\email{xupeng@imech.ac.cn}

\author[orcid=0000-0002-8174-0128]{Xilong Fan}
\affiliation{School of Physics Science And Technology, Wuhan University, No.299 Bayi Road, Wuhan, Hubei, China}
\email{xilong.fan@whu.edu.cn}

\correspondingauthor{Minghui Du, Xilong Fan}\email[show]{duminghui@imech.ac.cn, xilong.fan@whu.edu.cn}

% \collaboration{all}{The Terra Mater collaboration}

\begin{abstract}

The cosmic distance duality relation (CDDR), expressed as $d_L(z) = (1+z)^2 D_A(z)$, is a fundamental relation in modern cosmology. In this work, we apply a method to test the CDDR using simulated strongly lensed gravitational-wave (SLGW) signals from massive binary black holes (MBBH) as observed by proposed space-based detector networks. Our analysis is conducted under the point-mass lens model, considering the strong lensing scenario that produces two images. We generate 90 days of simulated SLGW data for 10 events based on the Population III stellar formation model, with source redshifts in the range $z_s \in [2,6]$ and lens redshifts in $z_L \in [0.2,1]$. The deviation of CDDR is parameterized by $\eta_1(z) = 1 + \eta_0 z$ and $\eta_2(z) = 1 + \eta_0 z/(1+z)$, and we incorporate the deviation parameter $\eta_0$ directly into the waveform model. Parameter estimation is performed within a Bayesian statistical framework, combining simulated data from both Taiji and LISA. For a single lensed event, the joint Taiji+LISA analysis improves the measurement precision of $\eta_0$ by roughly a factor of two compared with Taiji-only observations. By combining 10 simulated events, the population-level constraints on $\eta_0$,
quantified by the half width of the $95\%$ credible interval, reach approximately $2.61\times10^{-4}$ ($1.72\times10^{-4}$) for the $\eta_1(z)$ parameterization and $1.22\times10^{-3}$ ($6.86\times10^{-4}$) for $\eta_2(z)$ in the Taiji-only (Taiji+LISA) scenario, respectively. The inferred values of $\eta_0$ remain consistent with $\eta_0 = 0$ within the estimated uncertainties, with no statistically significant evidence for deviations from the CDDR at the achieved precision. These results demonstrate the significant advantage of joint space-based observations for high-precision tests of the CDDR.

\end{abstract}

% \keywords{\uat{Galaxies}{573} --- \uat{Cosmology}{343} --- \uat{High Energy astrophysics}{739} --- \uat{Interstellar medium}{847} --- \uat{Stellar astronomy}{1583} --- \uat{Solar physics}{1476}}
\keywords{\uat{Gravitational lensing}{670} --- \uat{Gravitational waves}{678} --- \uat{Bayesian statistics}{1900} --- \uat{Cosmology}{343}}


\section{Introduction} 

Modern cosmological studies rely critically on accurate measurements of astronomical distances, particularly the angular diameter distance $D_A(z)$ and the luminosity distance $d_L(z)$. While the redshift $z$ of a celestial object can be determined with high precision from its spectral features, reliable determinations of cosmological distances remain fundamentally important. In a general spacetime, $d_L(z)$ and $D_A(z)$ constitute the two primary distance measures that can be directly inferred from observations. Within the framework of a metric theory of gravity, under the assumptions that photons propagate along null geodesics and that their number is conserved, these two distance measures are not independent but are linked by $d_L(z) \equiv (1+z)^2 D_A(z)$, \citep{Etherington1933, Ellis2007}. 
Testing the validity of the CDDR provides an important consistency check of metric theories of gravity and the standard cosmological framework, and offers a probe of possible new physical or astrophysical processes \citep{Bassett2004PhRvD, Corasaniti2006MNRAS, Ellis2013PhRvD}. Any statistically significant deviation from this relation would imply a breakdown of one or more of its underlying assumptions, such as photon number conservation or photon propagation along null geodesics, and would therefore indicate physics beyond the standard cosmological model.
Moreover, the CDDR plays a pivotal role in a wide range of astronomical applications, including constraints on cosmic curvature derived from strong gravitational lensing systems \citep{Liu2020MNRAS, Xia2017ApJ, Qi2019MNRAS}, studies of the large-scale structure of galaxies and the remarkable uniformity of the cosmic microwave background temperature \citep{Planck2020AA}, and investigations of the geometric configuration, gas mass density distribution, and thermal structure of galaxy clusters \citep{Cao2011SCPMA, Cao2016MNRAS, Holanda2011AA}.
 
Testing the CDDR therefore provides a direct probe of one of the key relations in observational cosmology. Such tests require independent measurements of $D_A(z)$ and $d_L(z)$ at the same redshift, followed by comparison of their observed ratio with theoretical predictions. Numerous studies have investigated the CDDR using a variety of observational datasets \citep{Holanda2010ApJ,Li2011ApJL,Liang2013MNRAS,Lima2021JCAP,Wu2015PhRvD,Ruan2018ApJ,Liao2022ChPhL,Li2018MNRAS,Xu2022ApJ,Liao2016ApJ,Tang2023ChPhC,Liu2023PhLB,Lin2020ChPhC,Lin2021ChPhC,Arjona2021PhRvD}. Type Ia supernovae, as reliable standard candles, are widely used to measure $d_L(z)$, whereas $D_A(z)$ is typically inferred from galaxy clusters through the Sunyaev--Zeldovich effect and gas mass fraction measurements \citep{Holanda2010ApJ,Li2011ApJL,Liang2013MNRAS,Lima2021JCAP, Li2025PhRvD, Li2025arXiv}, baryon acoustic oscillations \citep{Wu2015PhRvD}, strong gravitational lensing systems \citep{Ruan2018ApJ,Liao2022ChPhL}, and the angular sizes of ultra-compact radio sources \citep{Li2018MNRAS}. For example, combining $D_A(z)$ from galaxy clusters with type Ia supernovae provides a direct test of the CDDR \citep{Holanda2010ApJ,Li2011ApJL,Liang2013MNRAS,Lima2021JCAP}. Xu et al.\ \citep{Xu2022ApJ} proposed a model-independent test of the CDDR by combining the latest five baryon acoustic oscillations measurements with the Pantheon type Ia supernovae sample, while Liao et al.\ \citep{Liao2016ApJ} introduced an alternative model-independent approach based on strong gravitational lensing systems and type Ia supernovae. However, the limited redshift coverage of type Ia supernovae renders strong gravitational lensing systems with source redshifts greater than $z \gtrsim 1.4$ unsuitable for direct CDDR tests, since no corresponding type Ia supernovae are available at the same redshift, resulting in a substantially reduced number of usable data pairs. To overcome this limitation, Tang et al.\ \citep{Tang2023ChPhC} reconstructed $d_L(z)$ from type Ia supernovae using deep learning up to the highest redshifts probed by strong gravitational lensing systems, and Liu et al.\ \citep{Liu2023PhLB} proposed a method combining ultra-compact radio measurements with the latest type Ia supernovae data and employing artificial neural networks to reconstruct the possible redshift evolution of the CDDR, enabling a fully model-independent test.

Traditional approaches to testing the CDDR suffer from several intrinsic limitations. First, distance measurements inferred from electromagnetic (EM) observations are susceptible to cosmic opacity effects, which can bias the determination of luminosity distances \citep{Avgoustidis2010JCAP,Liao2015PhRvD,Li2013PhRvD}. Second, these methods generally rely on the assumption of cosmic isotropy, as distances are inferred from different astrophysical objects distributed across distinct redshifts and sky directions, requiring interpolation procedures to combine heterogeneous measurements for the CDDR test \citep{Li2019MNRAS}. With the successful detection of gravitational waves (GWs) \citep{Abbott2024PhRvD,Abbott2019PhRvX,Abbott2021PhRvX,Abbott2021ApJ,Abbott2023PhRvX,LIGO2025arXiv}, astronomy and cosmology have entered the era of GW observations \citep{Abbott2016PhRvL,Abbott2017PhRvL}. In the near future, third-generation ground-based GW detectors will enable the detection of strongly lensed gravitational-wave (SLGW) signals, which are expected to provide a powerful and independent probe for testing the CDDR and mitigating the limitations discussed above \citep{Lin2020ChPhC,Lin2021ChPhC,Arjona2021PhRvD}. This advantage arises because GW propagation is insensitive to cosmic opacity, enabling a direct and unbiased measurement of the luminosity distance $d_L(z)$. Moreover, both the luminosity and angular-diameter distances can be inferred from the same lensed source through the combination of GW time-delay information and complementary EM lensing observables—such as the Einstein radius and the stellar velocity dispersion of the lens galaxy—thereby avoiding the need to combine distances from different sky locations and reducing the reliance on the cosmic isotropy assumption.

Future space-based GW observatories, including Taiji \citep{Hu2017NSRev}, TianQin \citep{Luo2016CQGra}, and the Laser Interferometer Space Antenna (LISA; \cite{Pau2017arXiv}), are expected to detect SLGW events at appreciable rates, thereby providing a realistic observational foundation for cosmological applications. For such systems, determining the source redshift requires the identification of EM counterparts. This requirement is physically well motivated, as MBBH evolving in gas-rich environments can power substantial EM emission through accretion onto the massive black hole across a broad range of wavelengths \citep{Ascoli2018ApJ}, enabling multi-messenger observations and host-galaxy identification. In addition, the lensing galaxies are typically luminous systems whose redshifts can be measured through existing photometric and spectroscopic galaxy surveys, allowing both the source and lens redshifts to be independently determined. Current observations further support the detectability of such EM signatures. For instance, the JWST has detected quasars at redshifts as high as $z=7.642$ \citep{Pacucci2022MNRAS}, demonstrating that luminous accreting massive black hole can be observed deep into the cosmic dawn. Provided that the GW sky-localization accuracy is sufficiently high, the association with EM counterparts becomes feasible \citep{Tamanini2016JCAP}. Under optimistic astrophysical scenarios, the GW detection rate from MBBH mergers is expected to reach hundreds to thousands of events, implying that the cumulative number of observable SLGW systems could be as large as $\gtrsim \mathcal{O}(5)$ over the operational lifetimes of Taiji and LISA \citep{Klein2016PhRvD,Diao2025arXiv,Ruan2021Resea,Wang2019PhRvD,Yuan2026ApJ}. The Taiji Data Challenge \citep{Du2025arXiv} further demonstrates the feasibility of realistic signal simulations and parameter-estimation pipelines for these sources.

Recently, \citet{Huang2025PDU} proposed a test of the CDDR using SLGW signals observed by TianQin. Owing to differences in interferometer arm lengths and noise budgets, space-based GW detectors are optimized for partially complementary frequency bands. In particular, TianQin achieves higher sensitivity at relatively high frequencies, whereas Taiji and LISA are designed to provide enhanced sensitivity at lower frequencies, which are especially relevant for MBBH systems. The similar sensitivity bands of Taiji and LISA also enable the formation of a detector network capable of coherently observing the same sources, leading to improved signal-to-noise ratios, parameter estimation accuracy, and sky-localization performance. These characteristics make such a network particularly suitable for SLGW-based cosmological studies. The potential of using SLGW signals observed by Taiji, LISA, and their joint network for CDDR tests has yet to be fully explored and represents a promising avenue for future research. It should be noted that these tests critically rely on complementary EM observations to provide independent redshift measurements for both the GW source and the intervening lens, making the combination of space-based GW detections and EM redshift information a key ingredient for precision cosmology.

Building on this motivation, we present the first joint Taiji--LISA analysis to constrain the CDDR using SLGW events within a Bayesian inference framework. By combining space-based GW observations with EM-identified redshifts for both the source and the lens, our approach enables a self-consistent multi-messenger test of the CDDR that is intrinsically free from cosmic opacity effects. We further evaluate the constraining power of this framework by forecasting, in a Bayesian context, the precision achievable on the CDDR from a population of ten multi-messenger SLGW events expected over the operational lifetimes of Taiji and LISA. These results highlight the strong potential of coordinated space-based GW observations for precision cosmology in the multi-messenger era.

The paper is organized as follows. Section~\ref{sec:lens} provides a brief overview of the point-mass lensing model used to describe strongly lensed GW signals. In Section~\ref{sec:bayes}, we present the Bayesian framework for inferring source and lens parameters, as well as the CDDR deviation parameter $\eta$. Section~\ref{sec:sim} describes the simulation of SLGW signals and the information required from both the source and lens, including the source luminosity distance, lens angular diameter distance, and redshifts, to perform a self-consistent CDDR test. Section~\ref{sec:res} presents the resulting parameter constraints, and Section~\ref{sec:con} summarizes the main findings and discusses possible extensions. Throughout this paper, we assume a flat $\Lambda$CDM cosmology with $\Omega_m=0.3111$, $\Omega_\Lambda=1-\Omega_m$, and $H_0=67.66\ {\rm km/s/Mpc}$ \citep{Planck2020AA}.

\section{Method} \label{sec:method}

\subsection{Point-Mass Model}\label{sec:lens}

\begin{figure}[h]
\centering 
\includegraphics[width=8.5cm]{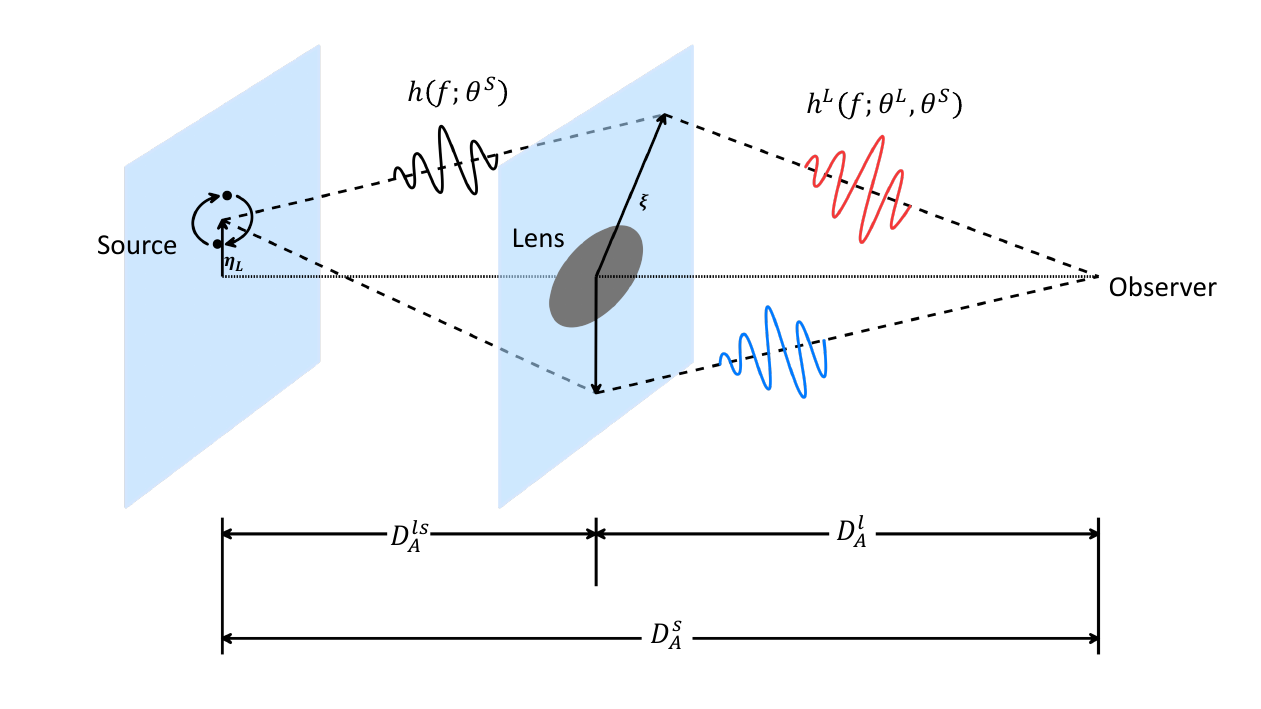}
\caption{Schematic illustration of GW lensing. The vectors $\eta_L$ and $\xi$ represent the positions on the source plane and the lens plane, respectively. The angular diameter distances $D_A^{l}$, $D_A^{s}$, and $D_A^{ls}$ are measured in the observer's rest frame.}
\label{fig:lens_show}
\end{figure}

For simplicity, we adopt a point mass model for black holes or stars as the lensing model for SLGWs. The basic picture of strong GW lensing is illustrated in Fig.~\ref{fig:lens_show}, where $D_A^l$, $D_A^s$, and $D_A^{ls}$ denote the angular diameter distances to the lens, to the source, and between the lens and the source, respectively.

In this framework, the diffraction of GWs can be modeled following \citet{Takahashi2003ApJ}. In the point mass model, the entire lensing mass is concentrated at a single point, and the corresponding mass density is expressed as
\begin{equation}
\rho_\mathrm{PM}(\bm r) = M_L \delta^3(\bm r),
\end{equation}
where $\delta^3(\bm r)$ is the three-dimensional Dirac delta function and $M_L$ is the lens mass. As the normalization constant, we adopt the Einstein radius

\begin{equation}
\xi_0 \equiv \left(\frac{4 G M_L D_A^l D_A^{ls}}{c^2 D_A^s}\right)^{1/2},
\end{equation} 
% in this expression, $D_L$, $D_S$, and $D_{LS}$ denoted the angular diameter distances from the observer to the lens, from the observer to the source, and from the lens to the source, respectively.
and the dimensionless deflection potential is
\begin{equation}
\psi(x) = \ln x.
\end{equation}

The amplification factor for a point mass lens is then given by
\begin{equation}
\begin{aligned}
F(w, y) &= \exp\left\{\frac{\pi w}{4} + i \frac{w}{2} \left[\ln\frac{w}{2} - 2\phi(y)\right]\right\} \\
& \quad \times \Gamma\left(1 - \frac{i w}{2}\right) \, _1F_1\left(\frac{i w}{2}, 1; \frac{i w y^2}{2}\right),
\label{eq:Fw}
\end{aligned}
\end{equation}
where $w = 8 \pi G M_L (1+z_L) f / c^3$ is the dimensionless frequency, $\phi(y) = (x_+ - y)^2/2 - \ln x_+$, $x_+ = (\sqrt{y^2 + 4} + y)/2$, $\Gamma$ is the Gamma function, and $_1F_1(a,b;z)$ is the confluent hypergeometric function (complex Kummer function). 

In the geometrical optics limit ($w \gg 1$), Eq.~(\ref{eq:Fw}) reduces to
\begin{equation}
F(w,y) = |\mu_+|^{1/2} - i \, |\mu_-|^{1/2} e^{2 \pi i f \Delta t_d},
\end{equation}
where the magnifications of the individual images are
\begin{equation}
\mu_\pm = \frac{1}{2} \pm \frac{y^2 + 2}{2 y \sqrt{y^2 + 4}},
\end{equation}
and the time delay between the two images is
\begin{equation}
\Delta t_d = \frac{4 G M_L (1+z_L)}{c^3} \left[ \frac{y \sqrt{y^2 + 4}}{2} + \ln \left( \frac{\sqrt{y^2 + 4} + y}{\sqrt{y^2 + 4} - y} \right) \right].
\end{equation}

Here, the dimensionless source position in the source plane is defined as
\begin{equation}
y = \frac{\eta_L c}{2 \sqrt{G M_L}} \sqrt{\frac{D_A^l}{D_A^s D_A^{ls}}},
\end{equation}
where $\eta_L$ denotes the transverse physical position of the GW source projected onto the source plane.
% where $\eta_L$ denotes the GW source position in the lens plane.

\subsection{Bayesian statistical framework}\label{sec:bayes}

We simulate GW signals from mergers of MBBH based on the Taiji and LISA detector configurations. Throughout this work, we exclusively employ the noise-orthogonal TDI-$A_2$ and $E_2$ channels for Bayesian inference \citep{Prince2002PhRvD}. Further technical details of the simulation setup can be found in \citet{Yuan2026ApJ}. The simulated unlensed GW signal in the frequency domain is constructed from Eqs.~(\ref{eq:un-d}) and (\ref{eq:tdi}):
\begin{equation}
d(f;\theta^S) = h(f;\theta^S) + n(f),
\label{eq:un-d}
\end{equation}
\begin{equation}
h(f;\theta^S) = h_{\tilde{A}_2,\tilde{E}_2}(f; \theta^S)
= \mathcal{T}_{A_2, E_2}(f, t_f)\, \tilde{h}(f;\theta^S),
\label{eq:tdi}
\end{equation}
where $\mathcal{T}_{A_2, E_2}^{\ell m}$ denotes the TDI transfer function, $h(f;\theta^S)$ represents the unlensed GW signal, $\tilde{h}(f;\theta^S)$ represent the GW waveform and $n(f)$ is the detector noise.

In this work, we neglect the effects of black hole spins. As a result, the source-parameter vector is simplified to
\begin{equation}
\theta^S = (\mathcal{M}_c, q, d_\mathrm{L}, t_c, \phi_c, \iota, \lambda, \beta, \psi),
\end{equation}
where $\mathcal{M}_c$ is the chirp mass, $q$ is the mass ratio, $d_\mathrm{L}$ is the luminosity distance, $t_c$ and $\phi_c$ are the coalescence time and phase, respectively, $\iota$ is the inclination angle, $(\lambda, \beta)$ denote the ecliptic longitude and latitude of the source, and $\psi$ is the polarization angle.

To test for deviations from the CDDR, we parameterize it as 
\begin{equation}
(1+z)^2 D_A / d_L = \eta(z),
\end{equation} 
and expand $\eta(z)$ using two approaches (see \citep{Yang2013ApJL, Holanda2016JCAP, Liao2019ApJ} for references):

\begin{enumerate}[label=\alph*.]
    \item Expansion in redshift $z$: $\eta_1(z) = 1 + \eta_0 z$;
    \item Expansion in scale factor $a = 1/(1+z)$: $\eta_2(z) = 1 + \eta_0 z/(1+z)$, which avoids divergences at high redshift.
\end{enumerate}

We adopt the standard distance--redshift relation \citep{Hogg1999astro} to reparameterize  the angular diameter distances $D_A^l$, $D_A^s$, and $D_A^{ls}$:
\begin{equation}
D_A(z) = \frac{1}{H_0 (1+z)} \int_0^{z} \frac{dz'}{E(z')},
\end{equation}
where 
\begin{equation}
E(z) = \sqrt{\Omega_M (1+z)^3 + \Omega_\Lambda}
\end{equation}
describes the background evolution of the Universe in a flat $\Lambda$CDM cosmology.

Finally, the luminosity distance appearing in the GW waveform is replaced by the CDDR-modified expression,
\begin{equation}
d_L = \frac{D_A^s (1+z_s)^2}{\eta(z_s)},
\label{eq:dl}
\end{equation}
where $z_s$ is the redshift of the GW source. In the geometrical-optics limit of a point mass lens model, the signal of SLGW can be expressed as
\begin{equation}
h^{L}(f;\theta^S) = F(w,y)h\left(f;\theta^S\right)
\label{eq:hwy}
\end{equation}

Based on the definitions introduced above, the luminosity distance $d_L$ is a function of the redshift and $\eta(z)$, the dimensionless frequency $w$ depends on the GW frequency, the lens mass $M_L$, and the lens redshift $z_L$, while $y$ is a function of $\eta_L$. We therefore reclassify the model parameters into three subsets,
$\theta^Z = (\mathcal{M}_c, q, z_s, t_c, \phi_c, \iota, \lambda, \beta, \psi)$, 
$\theta^L = (M_L, z_L, \eta_L)$ and $\theta^\eta=(\eta)$,
where $\theta^Z$ denotes the source parameters, $\theta^L$ represents the lens-related parameters and $\theta^\eta$ denotes the parameters used to test the CDDR.

Based on the above parameter classification, Eq.~(\ref{eq:hwy}) can be reformulated as
\begin{equation}
h^{L}(f;\theta^Z,\theta^L,\eta) = F(f;\theta^L)\, h(f;\theta^Z,\eta),
\label{eq:hL}
\end{equation}
where $F(f;\theta^L)$ is the lensing amplification factor and $h(f;\theta^Z,\eta)$ denotes the unlensed GW signal.

Consequently, the simulated observed data after lensing are given by
\begin{equation}
d^L(f) = h^{L}(f;\theta^Z,\theta^L,\eta) + n(f),
\end{equation}
with $n(f)$ representing the detector noise.

Within the Bayesian statistical framework, we perform a systematic analysis of parameter estimation for individual GW events observed by a single detector using the simulated lensed GW data. The corresponding Bayesian formulation for a single lensed event is
\begin{equation}
p(\theta^Z, \theta^L, \eta \mid d^L) 
= \frac{\mathcal{L}(d^L \mid \theta^Z, \theta^L, \eta)\, p(\theta^Z, \theta^L, \eta)}
       {\mathcal{Z}^L},
\label{eq:B-lens}
\end{equation}
where $\mathcal{L}$ is the likelihood function, $p$ denotes the prior distribution, and $\mathcal{Z}^L$ is the Bayesian evidence for the lensed model:
\begin{equation}
\mathcal{Z}^L = \int \mathcal{L}(d^L \mid \theta^Z, \theta^L, \eta)\, p(\theta^Z, \theta^L, \eta)\, 
\mathrm{d}\theta^Z\, \mathrm{d}\theta^L\, \mathrm{d}\eta.
\end{equation}

% For joint parameter estimation and testing of the CDDR, we consider the combined observations from both the Taiji and LISA detectors. Denoting the lensed GW data from Taiji and LISA as $d^L_\mathrm{Taiji}$ and $d^L_\mathrm{LISA}$, respectively, the joint posterior for the source parameters $\theta^Z$, lens parameters $\theta^L$, and the CDDR parameter $\eta$ can be expressed as
For joint parameter estimation and testing of the CDDR, we consider the combined observations from both the Taiji and LISA detectors. Denoting the lensed GW data from Taiji and LISA as $d^L_{\mathrm{Taiji}}$ and $d^L_{\mathrm{LISA}}$, respectively, we assume that the instrumental noise in the two detectors is statistically independent, owing to their distinct orbital configurations and independent measurement systems. Under this assumption, the joint likelihood factorizes into the product of the individual likelihoods. The joint posterior distribution for the source parameters $\theta^Z$, lens parameters $\theta^L$, and the CDDR parameter $\eta$ can therefore be written as
\begin{equation}
\begin{aligned}
p(\theta^Z, \theta^L, \eta \mid d^L_\mathrm{Taiji}, d^L_\mathrm{LISA})
&\propto 
\mathcal{L}_\mathrm{Taiji}(d^L_\mathrm{Taiji} \mid \theta^Z, \theta^L, \eta) \\
&\quad \times 
\mathcal{L}_\mathrm{LISA}(d^L_\mathrm{LISA} \mid \theta^Z, \theta^L, \eta) \\
&\quad \times 
p(\theta^Z, \theta^L, \eta),
\end{aligned}
\label{eq:B-joint}
\end{equation}
where $\mathcal{L}_\mathrm{Taiji}$ and $\mathcal{L}_\mathrm{LISA}$ are the likelihood functions for the respective detectors. This framework enables a combined analysis of GW events observed by multiple space-based detectors and provides a more stringent test of the CDDR.

% Finally, we simulate ten cases of joint Taiji--LISA observations and perform parameter estimation for each. By analyzing these ten cases, we evaluate the improvement in constraining the CDDR parameter $\eta$ achieved by combining multiple GW events observed jointly by the two detectors. The joint posterior for $\eta$ across $n$ events is then given by
Finally, we simulate ten joint Taiji--LISA observations and perform parameter estimation for each event individually. Since the events are astrophysically distinct and their detector noise realizations are independent, the corresponding likelihoods are statistically independent. By combining these ten events, we quantify the improvement in constraining the CDDR parameter $\eta$ enabled by multiple jointly observed GW signals. Under the independence assumption, the joint posterior for $\eta$ from $n$ events can be written as
\begin{equation}
p(\eta) = \prod_{i=1}^{n} 
\int p(\theta^Z_i, \theta^L_i, \eta \mid d^L_{\mathrm{Taiji},i}, d^L_{\mathrm{LISA},i}) 
\, \mathrm{d}\theta^Z_i \, \mathrm{d}\theta^L_i.
\end{equation}
This approach allows us to quantify the statistical gain in constraining $\eta$ when combining multiple lensed GW events in a multi-detector framework. 
In addition, complementary EM observations provide independent measurements of the source and lens redshifts, as well as accurate sky-localization information through host-galaxy identification. 
These quantities can therefore be treated as externally constrained. Accordingly, in the subsequent parameter-estimation analysis, we fix the redshifts of both the lens and the source and the sky-position parameters, and infer the remaining GW and lensing parameters together with the CDDR deviation parameter from the data.

\section{Data analysis}

\subsection{Data simulation and waveform analysis}\label{sec:sim}

Under the point mass lens model, we simulate SLGW signals from MBBH according to Eq.~(\ref{eq:hL}). The unlensed GW waveforms are generated using the IMRPhenomD model \citep{Husa2016PhRvD, Khan2016PhRvD}, which accurately describes the dominant $\ell = 2$, $|m| = 2$ modes of aligned-spin binary systems. In this study, we consider only the dominant mode and neglect higher-order modes, as this approximation is sufficient to capture the main features of SLGW signals. The data simulation for the Taiji mission has been described in detail in the Appendix of \citet{Yuan2026ApJ}. The simulation procedure for the LISA mission is analogous, with the only modification being the replacement of the noise sensitivity curve and detector orbit. For the LISA sensitivity curve and related instrumental parameters, we refer the reader to \citet{Robson2019CQGra}.

During the data simulation, without loss of generality, we set the starting point to day 60 of the  mission orbit and simulate a total duration of 90 days.
To ensure that the simulated data cover the inspiral, merger, and ringdown phases of the MBBH system, the merger time is placed on the day before the end of the segment, i.e., day 149. In addition, when analyzing the case of binary systems with component masses of $10^5 M_\odot$, the sampling frequency of the space-based GW detector is set to 0.3, while for systems with component masses greater than $10^5 M_\odot$, the sampling frequency is set to 0.03. This configuration ensures both waveform accuracy and improved computational efficiency.

To investigate the performance of multi-event joint observations with Taiji and LISA, we simulate a set of ten multi-messenger SLGW events. The simulation is constructed to ensure both physical consistency and observational feasibility. For a controlled comparison between the two CDDR parameterizations, identical source, lens, and noise realizations are adopted for the $\eta_1(z)$ and $\eta_2(z)$ cases, such that any differences in the inferred constraints arise solely from the choice of parameterization rather than variations in the underlying event properties.

For the source population, we generate MBBH systems following a Population~III \citep{Vikaeus2022MNRAS,Diao2025arXiv} star formation scenario. The total mass of each binary is drawn from the range $10^{4}\,M_\odot \leq M_{\rm tot} \leq 10^{7}\,M_\odot$, which covers the mass interval to which space-based GW detectors are most sensitive. We further impose a mass-ratio constraint of $q \geq 0.75$, motivated by theoretical predictions that Population~III binaries preferentially form nearly equal-mass systems. The source redshift is restricted to $z_{\rm s} \in [2,6]$. This choice ensures that the GW signals are detectable by both Taiji and LISA, while also allowing the possibility of detectable EM emission from massive black holes evolving in gas-rich environments, thereby enabling multi-messenger identification with current or near-future EM observatories \citep{Ascoli2018ApJ}.

% Since the coalescence time, coalescence phase, orbital inclination, polarization angle, and sky location are not directly relevant to the CDDR constraints considered in this work, we fix these parameters to representative values for all ten simulated events to simplify the analysis and reduce the computational cost. This treatment is adopted purely for practical purposes, while our inference framework remains applicable to the full parameter space. Specifically, we adopt
Since the coalescence time, coalescence phase, orbital inclination, polarization angle, and sky location have only a weak influence on the CDDR constraints considered here, we hold these parameters fixed at representative values for all ten simulated events for computational convenience. 
These variables primarily act as geometric or extrinsic degrees of freedom and exhibit minimal correlation with the cosmological parameter $\eta$. 
Accordingly, we condition the analysis on fixed values for these quantities, while the inference framework itself remains fully general and directly extendable to the complete parameter space. 
Specifically, we adopt
\[
(t_c, \phi_c, \lambda, \beta, \iota, \psi)
= (149~\mathrm{day}, 4.32, 4.25, -0.15, 0.7, 0.6),
\]
which enables us to isolate and more clearly quantify the impact of lensing and cosmological parameters in the joint Taiji--LISA analysis.

For the lens population, we assume lens redshifts in the range $z_{\rm L} \in [0.2,1]$, corresponding to typical galaxy-scale lenses at intermediate cosmological distances. The lens mass is sampled uniformly in logarithmic space between $10^{9}\,M_\odot$ and $10^{12}\,M_\odot$, encompassing the characteristic mass range of galaxy and massive galaxy lenses responsible for strong gravitational lensing. Finally, to ensure that each simulated system produces at least two lensed images, we uniformly sample the dimensionless lensing parameter $\eta_{\rm L}$ within the interval $[1,5]$. This selection effectively excludes single-image configurations and guarantees that the resulting time delays and magnification factors fall within a regime that can be resolved and analyzed in joint Taiji--LISA observations.

% To begin with, we compare the differences between lensed and unlensed waveforms. The characteristic strain is selected as the metric for this comparison, since the area enclosed between the characteristic strain curve and the detector's noise curve corresponds to the commonly used SNR. This provides a more intuitive means of evaluating how lensing affects the waveform's detectability. The characteristic strain of the signal ($h_c$) and the noise ($h_n$) are defined in Eq. (\ref{eq:hc}) and Eq. (\ref{eq:hn}) \cite{Moore2015CQGra},
% \begin{equation}
% h_c(f)^2 = 4f^2|h(f;\theta^S)|^2,
% \label{eq:hc}
% \end{equation}
% \begin{equation}
% h_n(f)^2 = fS_n(f),
% \label{eq:hn}
% \end{equation}
% where $S_n(f)$ is one-sided noise power spectrum density (PSD). 

\subsection{Results}\label{sec:res}

Among the ten simulated events, we select one representative case to illustrate 
the constraints on the parameter $\eta_0$ obtained under the two $\eta(z)$ 
parameterizations introduced in the previous section.

The source parameters of the selected event are 
$M_{\rm tot} = 1.54 \times 10^{6}\,M_\odot$, 
$q = 0.85$.
% $t_c = 149~{\rm day}$, 
% $\phi_c = 4.32$, 
% $\iota = 0.7$, 
% and $\psi = 0.6$. 
The corresponding lens parameters are 
$\eta_{\rm L} = 3.0$, 
$M_{\rm L} = 2.7 \times 10^{10}\,M_\odot$, 
and $z_{\rm L} = 0.8$. 
For this event, we adopt $\eta_0 = 0$ as the injected value.

\begin{figure*}
\centering
\includegraphics[width=0.85\textwidth]{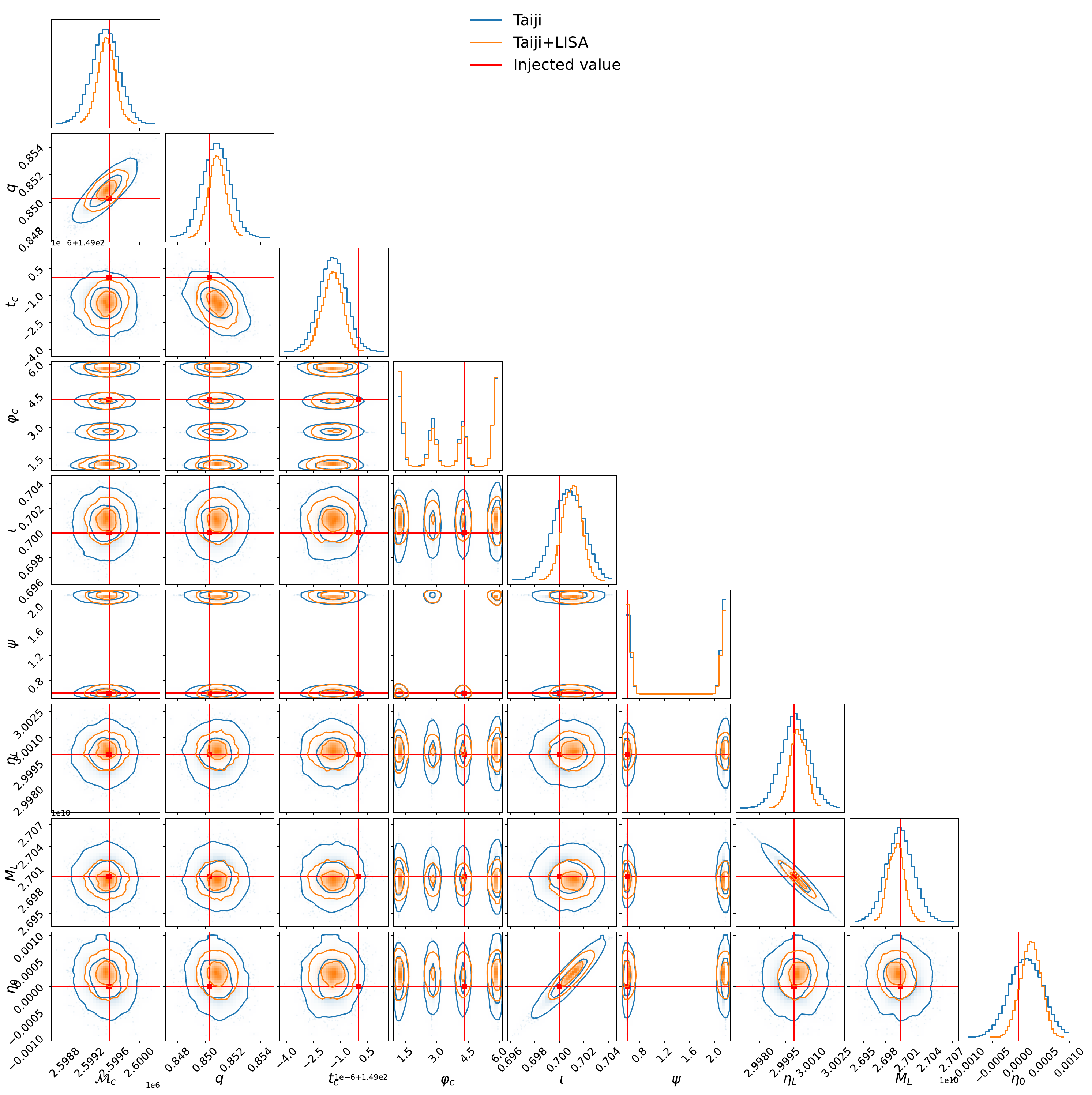}
\caption{
Posterior distributions of the source, lensing, and cosmological parameters for a representative SLGW event, assuming a parameterized deviation from the CDDR of the form $\eta_1(z) = 1 + \eta_0 z$. Blue contours denote the constraints obtained with Taiji alone, while orange contours correspond to the joint Taiji--LISA analysis. The inner and outer contours enclose the 50\% and 95\% credible regions, respectively. The red vertical and horizontal lines indicate the injected values of the parameters. The inferred posterior of $\eta_0$ remains compatible with $\eta_0 = 0$ within the quoted uncertainties, indicating no statistically significant evidence for a deviation from the CDDR in this event.}
\label{fig:case1}
\end{figure*}

\begin{figure*}
\centering
\includegraphics[width=0.85\textwidth]{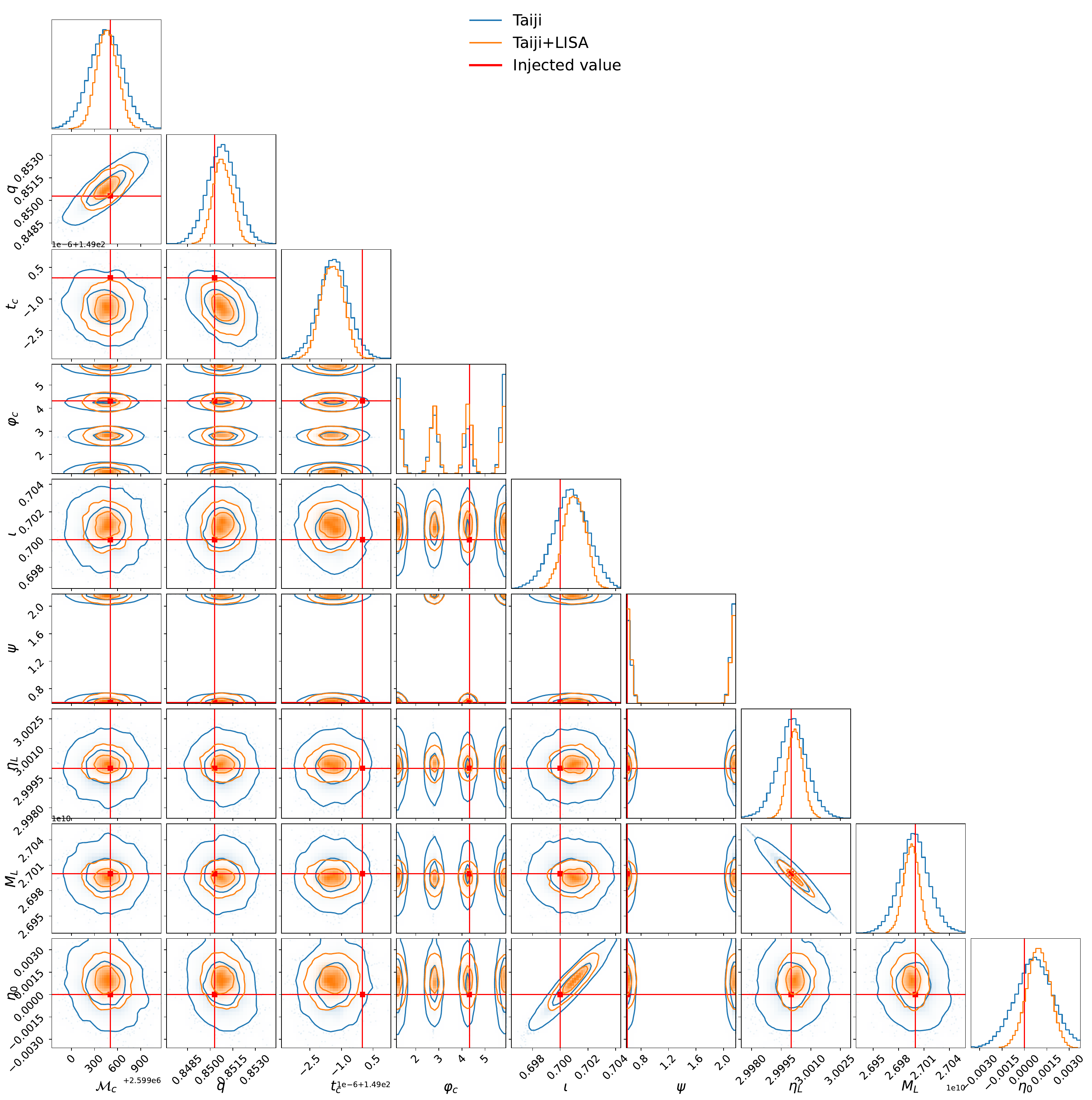}
\caption{
Same as Fig.~\ref{fig:case1}, but assuming an alternative parameterization of the deviation from the CDDR,
$\eta_2(z) = 1 + \eta_0 z/(1+z)$.
The inferred posterior of $\eta_0$ is also consistent with $\eta_0 = 0$, indicating no statistically significant deviation from the CDDR.}
\label{fig:case2}
\end{figure*}

Using the Bayesian posterior defined in Eq.~(\ref{eq:B-lens}) for a single-detector analysis and in Eq.~(\ref{eq:B-joint}) for the joint Taiji--LISA analysis, we perform parameter estimation for this event. 
The resulting posterior constraints obtained under the two $\eta(z)$ parameterizations are shown in Figures~\ref{fig:case1} and~\ref{fig:case2}, respectively.

Figure~\ref{fig:case1} quantitatively illustrates the differences in parameter
estimation performance between the Taiji single-detector observation and the joint Taiji--LISA observation, assuming the CDDR parameterization $\eta_1(z) = 1 + \eta_0 z$.

For the GW source parameters, the joint Taiji--LISA analysis consistently achieves higher measurement precision across nearly all relevant dimensions. Compared with the Taiji-only case, the one-dimensional posterior distributions of
the chirp mass $\mathcal{M}_c$ and the mass ratio $q$ are significantly narrower, with their
$95\%$ credible intervals substantially reduced.
This improvement reflects the enhanced signal-to-noise ratio and the partial breaking
of parameter degeneracies enabled by the multi-detector configuration. The posterior distributions of the coalescence time $t_c$ are also more tightly constrained, and the correlations observed in several two-dimensional posteriors are notably weakened. 
Overall, these results indicate that the joint Taiji+LISA analysis, benefiting from the different detector response functions and complementary sensitivity to the GW source, reduces correlations among parameters and improves the accuracy of source parameter estimation.

We find that for the coalescence phase $\phi_c$ and the polarization angle $\psi$,
their intrinsic degeneracy is not fully resolved in the joint Taiji--LISA analysis.
Among the ten analyzed events, the posterior distributions of $\phi_c$ and $\psi$
remain multimodal even in the joint analysis.
This behavior arises from an extended degeneracy along lines of constant
$\phi_c + \psi$ and $\phi_c - \psi$, which leads to mirror solutions in the joint
$(\phi_c, \psi)$ parameter space.
Therefore, while the inclusion of LISA significantly improves the overall parameter
constraints, it is insufficient by itself to break this specific degeneracy.
Incorporating higher-order gravitational-wave modes in the waveform model is expected
to effectively lift this degeneracy and suppress the resulting multimodality, as demonstrated in detailed studies such as \cite{Marsat2021PhRvD}.

The improvement is even more pronounced for the lensing parameters.
The posterior distributions of the lens mass $M_{\rm L}$ and the dimensionless
lensing parameter $\eta_{\rm L}$ obtained from the joint Taiji--LISA analysis are
significantly narrower than those from the Taiji-only analysis, with uncertainties
reduced by nearly an order of magnitude.
In addition, the correlations between $M_{\rm L}$ and the source mass parameters
are substantially suppressed, demonstrating the enhanced capability of the joint
observation to disentangle the intrinsic GW amplitude from lensing magnification
effects. 

Among the cosmological parameters, the CDDR deviation parameter $\eta_0$ shows the most substantial improvement when LISA observations are incorporated. The Taiji-only analysis yields $\eta_0 = -1.6\times10^{-4}{}^{+5.9\times10^{-4}}_{-6.2\times10^{-4}}$, while the joint Taiji--LISA analysis gives $\eta_0 = -2.5\times10^{-4}{}^{+3.6\times10^{-4}}_{-3.9\times10^{-4}}$, where all quoted uncertainties correspond to the 95\% confidence interval. The reduced uncertainty demonstrates the enhanced constraining power of the multi-detector configuration. 
In both cases, the inferred values of $\eta_0$ are consistent with $\eta_0 = 0$ within the quoted uncertainties, indicating that no statistically significant deviation from the CDDR is detected at the achieved level of precision. This provides an independent test of the CDDR in the redshift range probed by these SLGW events.
The inclusion of LISA reduces the width of the credible interval by approximately a factor of two, indicating a substantial enhancement in the precision of the
$\eta_0$ measurement.

In addition to the improved one-dimensional constraint, the joint analysis also significantly suppresses the correlations between $\eta_0$ and the lensing parameters, particularly the lens mass $M_{\rm L}$ and the dimensionless lensing parameter $\eta_{\rm L}$. This reduction in parameter degeneracies demonstrates that multi-detector observations are effective in disentangling cosmological effects encoded in $\eta_0$ from lensing-induced systematics that are inherent in single-detector measurements.

Figure~\ref{fig:case2} shows the quantitative comparison of parameter estimation between the Taiji single-detector observation and the joint Taiji--LISA observation, assuming an alternative parameterization of the CDDR, $\eta_2(z) = 1 + \eta_0 z / (1 + z)$.

For the source and lensing parameters, the results obtained under the $\eta_2(z)$ parameterization are fully consistent with those shown in Figure~\ref{fig:case1}. In particular, compared with the Taiji-only analysis, the joint Taiji--LISA observation yields systematically tighter posterior distributions for both the GW source parameters and the lensing parameters. The corresponding parameter correlations are also noticeably weakened, confirming that the inclusion of LISA data enhances the overall parameter estimation accuracy and improves the disentanglement between intrinsic source properties and lensing-induced effects.

For the cosmological parameter $\eta_0$, the Taiji-only analysis under the nonlinear CDDR parameterization $\eta_2(z)$ yields $\eta_0 = -6.5\times10^{-4}{}^{+2.4\times10^{-3}}_{-2.5\times10^{-3}}$ and $-9.5\times10^{-4}{}^{+1.4\times10^{-3}}_{-1.6\times10^{-3}}$, respectively (95\% confidence intervals). As in the case of $\eta_1(z)$, the inclusion of LISA leads to a clear reduction in the width of the credible interval and suppresses correlations between $\eta_0$ and the lensing parameters, demonstrating the robustness of the multi-detector improvement across different CDDR parameterizations.
In both analyses, the inferred values of $\eta_0$ are consistent with $\eta_0 = 0$ within the quoted uncertainties, indicating no statistically significant evidence for a deviation from the CDDR under this alternative parameterization.

Nevertheless, a direct comparison with Figure~\ref{fig:case1} shows that the overall uncertainty on $\eta_0$ obtained with $\eta_2(z)$ is significantly larger than that derived from the linear parameterization. This behavior is expected and originates from the intrinsic form of $\eta_2(z)$, in which the deviation term $\eta_0 z/(1+z)$ saturates at high redshift. As a result, high-redshift sources contribute less effectively to constraining $\eta_0$ in this parameterization, thereby reducing the overall sensitivity of
the analysis. This explains why, despite the improved performance of the joint Taiji--LISA observation, the constraints on $\eta_0$ in Figure~\ref{fig:case2} remain weaker than those obtained under $\eta_1(z)$ in Figure~\ref{fig:case1}.

Having established the parameter estimation results for individual lensed events, we next present the population-level constraints obtained by progressively combining multiple events. This analysis quantifies the improvement in the determination of the deviation parameter $\eta_0$ as the event sample increases.

\begin{figure}[htbp]
\centering
\includegraphics[width=0.47\textwidth]{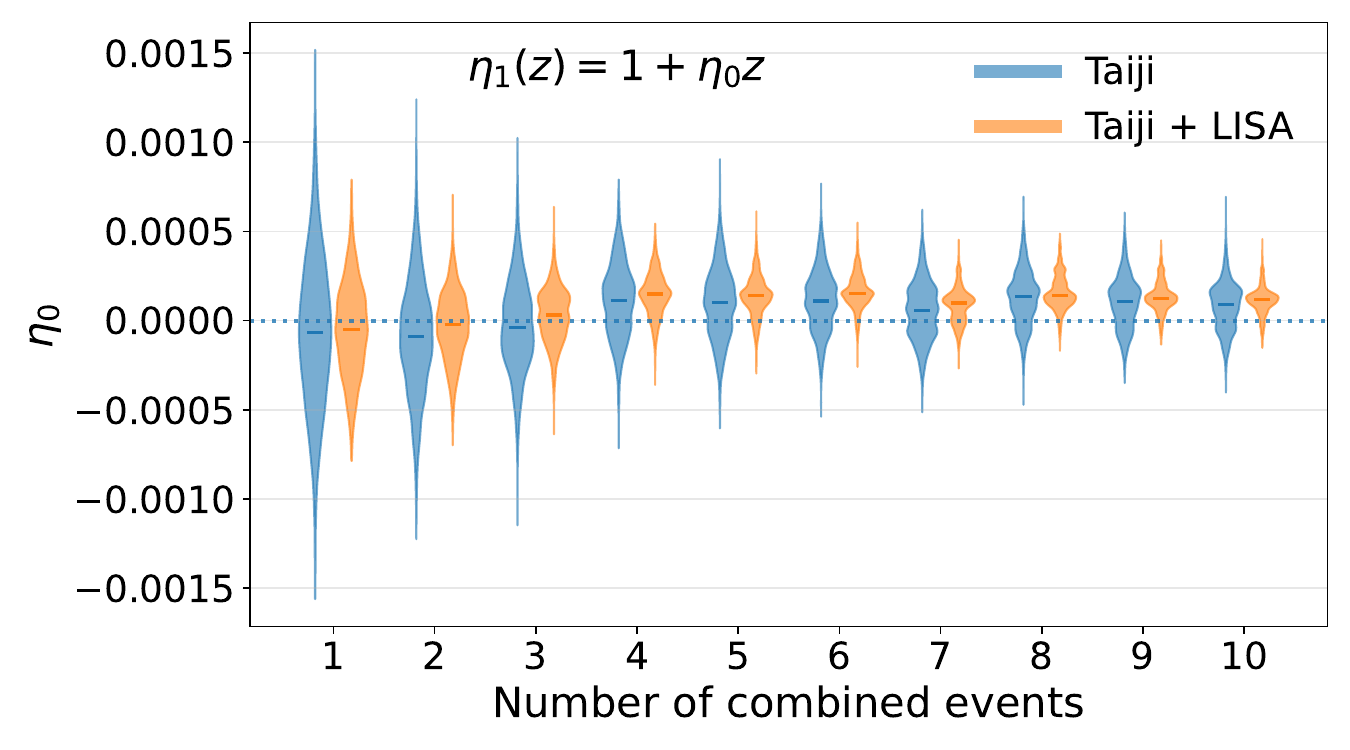}
\includegraphics[width=0.47\textwidth]{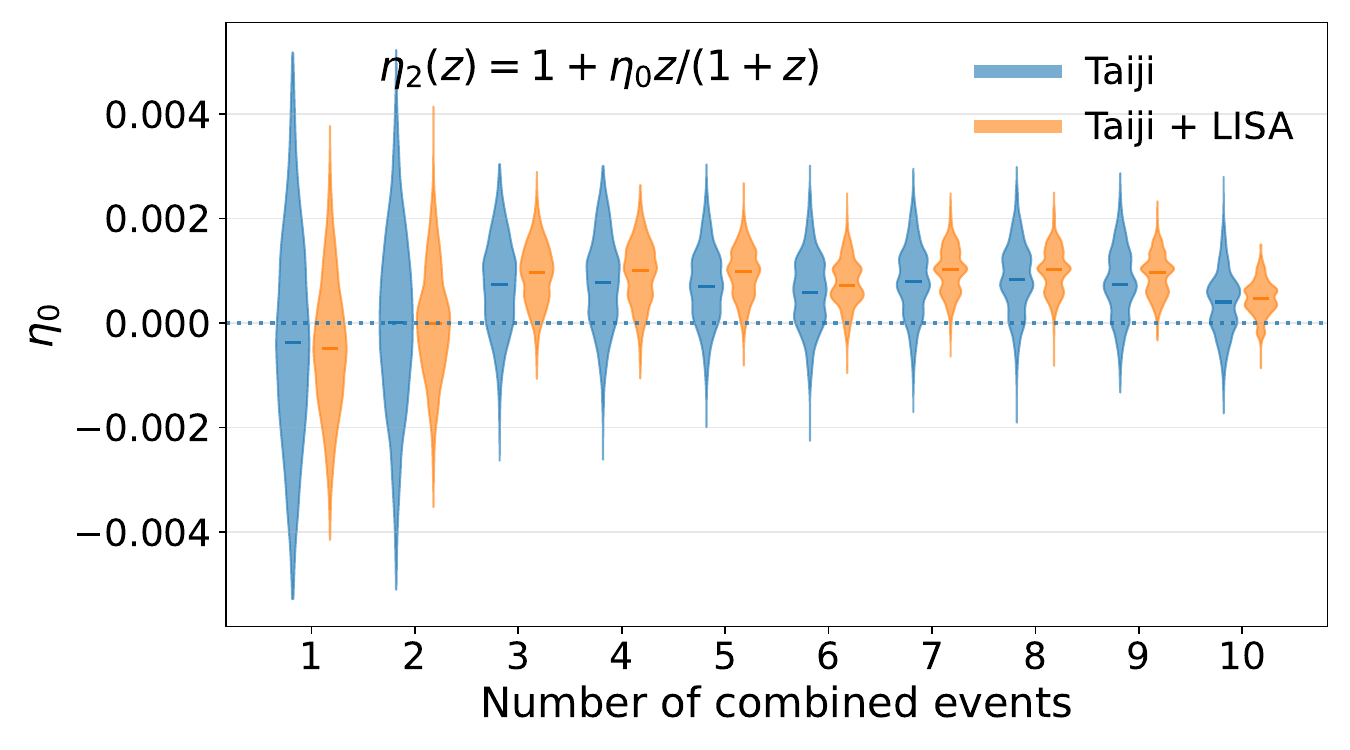}
\caption{
Population-level constraints on the deviation parameter $\eta_0$ obtained by progressively combining $10$ lensed events.
Top panel: results assuming the parametrization $\eta_1(z) = 1 + \eta_0 z$.
Bottom panel: results assuming $\eta_1(z) = 1 + \eta_0 z/(1+z)$.
Blue and orange violins correspond to Taiji-only and Taiji+LISA observations, respectively.
The horizontal dotted line indicates the fiducial value $\eta_0 = 0$.
In both case, the population posteriors remain compatible with $\eta_0 = 0$, indicating no evidence for a violation of the CDDR.}
\label{fig:eta_population}
\end{figure}

Figure~\ref{fig:eta_population} displays the population posteriors of $\eta_0$ inferred from the sequential combination of up to ten lensed events.  The top panel adopts a linear redshift dependence, 
$\eta_1(z) = 1 + \eta_0 z,$
while the bottom panel shows the results for
$\eta_2(z) = 1 + {\eta_0 z}/(1+z).$
For each number of combined events, the violin plots represent the reconstructed population posterior of $\eta_0$, obtained by multiplying the posteriors from individual events.

% For both parametrizations, the population posteriors remain centered on the fiducial value $\eta_0 = 0$, indicating no evidence for deviations from the distance–redshift duality relation. As additional events are included, the posterior widths decrease monotonically, reflecting the cumulative improvement in constraining power from the population analysis.
For both parametrizations, the population posteriors remain compatible with the fiducial value $\eta_0 = 0$, showing no statistically significant evidence for deviations from the distance--redshift duality relation. Furthermore, as additional events are incorporated into the analysis, the posterior widths decrease progressively, reflecting the expected cumulative improvement in constraining power achieved through the population analysis.

After combining ten events, we find that for the linear parameterization $\eta_1(z) = 1 + \eta_0 z$, the $95\%$ credible uncertainty on $\eta_0$ is $2.61\times10^{-4}$ using Taiji-only data, and is reduced to $1.72\times10^{-4}$ when jointly analyzing Taiji and LISA observations. For the nonlinear parameterization $\eta_2(z) = 1 + \eta_0 z/(1+z)$, the corresponding $95\%$ credible uncertainties are $1.22\times10^{-3}$ for Taiji alone and $6.86\times10^{-4}$ for the combined Taiji+LISA scenario. Comparing the two parameterizations, the constraints obtained with $\eta_1(z)$ are consistently tighter than those from $\eta_2(z)$, reflecting the reduced sensitivity of the nonlinear form to high-redshift sources, where the deviation term $\eta_0 z/(1+z)$ saturates.

Comparing the two observational configurations, the Taiji+LISA scenario consistently yields tighter population-level constraints on $\eta_0$ than the Taiji-only case. This improvement originates from the enhanced parameter estimation accuracy at the individual-event level when LISA data are included, which propagates coherently through the population combination. Overall, these results demonstrate the clear advantage of joint space-based GW observations for population studies of strongly lensed events and for precision tests of the cosmic distance duality relation.

\section{Conclusion}\label{sec:con}

In this work, we analyze the GW signals from MBBH expected in the Taiji and LISA frequency bands under the point-mass lens model, focusing on the impact of strong lensing that produces two distinct images on the observations. This setup allows us to test the CDDR in an opacity-free method. To this end, we construct 10 simulated events based on the Population III stellar formation model and generate 90 days of synthetic GW data for both Taiji and LISA. The source redshifts are distributed in the range $z_s \in [2,6]$, while the lens redshifts lie within $z_L \in [0.2,1]$. We perform parameter estimation using Taiji-only data as well as joint Taiji+LISA observations to investigate the constraints on the CDDR deviation parameter $\eta_0$. Furthermore, we assess the combined constraining power of the 10 simulated events, quantifying the potential precision with which $\eta_0$ can be measured.

For individual lensed events, as shown in Figures~\ref{fig:case1} and~\ref{fig:case2}, the joint Taiji+LISA analysis consistently achieves higher precision across both the source and lensing parameters compared with the Taiji-only configuration. In particular, the one-dimensional posterior distributions of the source chirp mass $\mathcal{M}_c$, mass ratio $q$, and coalescence time $t_c$ are significantly narrower, and degeneracies visible in the two-dimensional posteriors are substantially mitigated. Similarly, the lens mass $M_{\rm L}$ and the dimensionless lensing parameter $\eta_{\rm L}$ are more tightly constrained, with uncertainties reduced by nearly an order of magnitude in the joint analysis. For the CDDR deviation parameter $\eta_0$, under the $\eta_1(z)$ parameterization, the Taiji-only analysis yields $\eta_0 = -1.6\times10^{-4}{}^{+5.9\times10^{-4}}_{-6.2\times10^{-4}}$, while the joint Taiji+LISA analysis gives $\eta_0 = -2.5\times10^{-4}{}^{+3.6\times10^{-4}}_{-3.9\times10^{-4}}$. In both cases, the inferred values of $\eta_0$ remain consistent with $\eta_0 = 0$ within the quoted uncertainties, indicating no evidence for deviations from the CDDR at the level of precision achieved. Under the $\eta_2(z)$ parameterization, the corresponding constraints are $\eta_0 = -6.5\times10^{-4}{}^{+2.4\times10^{-3}}_{-2.5\times10^{-3}}$ and $\eta_0 = -9.5\times10^{-4}{}^{+1.4\times10^{-3}}_{-1.6\times10^{-3}}$, respectively. All quoted uncertainties correspond to the 95\% confidence intervals. In both parameterizations, the inclusion of LISA reduces the credible-interval width and suppresses correlations between $\eta_0$ and the lensing parameters, demonstrating the enhanced ability of the multi-detector configuration to disentangle cosmological effects from lensing-induced systematics.

% Extending the analysis to the population level, we sequentially combine up to ten simulated strongly lensed events to quantify the cumulative improvement in the constraint on the CDDR deviation parameter $\eta_0$. Figure~\ref{fig:eta_population} presents the resulting population posteriors for the two adopted parameterizations of the CDDR. For the linear parameterization $\eta_1(z) = 1 + \eta_0 z$, the $95\%$ credible uncertainty on $\eta_0$ reaches $2.61\times10^{-4}$ when using Taiji-only data and improves to $1.72\times10^{-4}$ with the combined Taiji+LISA observations. For the nonlinear parameterization $\eta_2(z) = 1 + \eta_0 z/(1+z)$, the corresponding uncertainties are $1.22\times10^{-3}$ and $6.86\times10^{-4}$, respectively.

Extending the analysis to the population level, we sequentially combine up to ten simulated strongly lensed events to quantify the cumulative improvement in the constraint on the CDDR deviation parameter $\eta_0$. Figure~\ref{fig:eta_population} presents the resulting population posteriors for the two adopted parameterizations of the CDDR. For the linear parameterization $\eta_1(z) = 1 + \eta_0 z$, the $95\%$ credible uncertainty on $\eta_0$ reaches $2.61\times10^{-4}$ when using Taiji-only data and improves to $1.72\times10^{-4}$ with the combined Taiji+LISA observations. For the nonlinear parameterization $\eta_2(z) = 1 + \eta_0 z/(1+z)$, the corresponding uncertainties are $1.22\times10^{-3}$ and $6.86\times10^{-4}$, respectively. In both cases, the population posteriors remain compatible with $\eta_0 = 0$, providing further support for the validity of the CDDR across the redshift range probed by these events.

A direct comparison between the two parameterizations shows that $\eta_1(z)$ consistently yields tighter constraints than $\eta_2(z)$, reflecting the reduced sensitivity of the latter to high-redshift sources due to the saturation of the deviation term at large $z$. Moreover, for both parameterizations, the Taiji+LISA configuration systematically outperforms the Taiji-only case, highlighting the crucial role of multi-detector synergy in enhancing individual-event parameter estimation and coherently propagating these gains to the population level. Overall, our results not only demonstrate the strong potential of multi-space-based GW detector networks for precision tests of the CDDR, but also provide an independent confirmation of the relation's validity under both linear and nonlinear parameterizations.

Finally, we emphasize that our analysis is conducted solely within the point mass lens model and that the simulated data used in this study are based on idealized assumptions. Specifically, we consider only GW signals from massive binary black holes and assume the presence of a single lens along the line of sight, neglecting the potential effects of multiple lenses or overlapping sources. In realistic Taiji and LISA observations, additional GW signals from other populations, such as double white dwarfs, would also be present and may contribute to the overall data. Future work will aim to extend our analysis to lensing signals embedded in more realistic observational data, including multiple lenses and overlapping sources, thereby providing a closer representation of actual space-based GW measurements.

%% Please use the acknowledgment and contribution environments. This will 
%% be anonomyized when the "anonymous" style option is used. 
\begin{acknowledgments}
We thank Shun-Jia Huang and Huan Zhou for helpful discussions and Qing Diao for generating the population parameters. This work is supported by National Key Research and Development Program of China grant Nos. 2024YFC2207300 and 2021YFC2201903. P.X. is supported by the International Partnership Program of the Chinese Academy of Sciences, grant No. 025GJHZ2023106GC.
\end{acknowledgments}

\begin{contribution}
%%This section gives authors the space to recognize author contributions. The text inside this environment is NOT counted towards the total word quanta. At a minimum, manuscripts are expected to include this text:

Yong Yuan: Conceptualization, Methodology, Software, Investigation, Formal analysis, Writing – original draft.  
Ming-Hui Du: Software, Writing – review \& editing.  
Ben-Yang Zhu: Software, Formal analysis.  
Xin-Yi Lin: Validation, Methodology. 
Wen-Fan Feng: Wrting - review.
Peng Xu: Supervision, Writing – review \& editing.  
Xi-Long Fan: Supervision, Methodology, Writing – review \& editing.  

%% But authors are expected to provide more specific details, e.g. 
%%
%%SC was responsible for writing and submitting the manuscript.
%%WWM came up with the initial research concept and edited the manuscript.
%%OTS obtained the funding and edited the manuscript.
%%EBF provided the formal analysis and validation. He also edited the manuscript.
%%GEH Supervised the undergraduates, wrote the software and administers the project github and Zenodo repositories.
%%
%% Authors can use the Contributor Role Taxonomy (CRediT) at
%% for ideas on how write a good statement tailored to their needs.

\end{contribution}

\bibliography{sample701}{}
\bibliographystyle{aasjournalv7}



\end{document}